# Flexible Hybrid Graphene / a-Si:H Multispectral Photodetectors

D. S. Schneider[1,*], A. Bablich[1,*], M. C. Lemme[1,2]

[1] University of Siegen, School of Science and Technology, Hölderlinstr. 3, Siegen, 57076, Germany

[2] RWTH Aachen University, Faculty of Electrical Engineering and Information Technology, Otto-Blumenthal-Str. 25, 52074 Aachen, Germany

[*] these authors contributed equally

E-Mail: max.lemme@rwth-aachen.de, Phone: +49 / 241 8867-200, Fax: +49 / 241 8867-560

## Abstract

We report on the integration of large area CVD grown single- and bilayer graphene transparent conductive electrodes (TCEs) on amorphous silicon multispectral photodetectors. The broadband transmission of graphene results in 440% enhancement of the detectors' spectral response in the ultraviolet (UV) region at $\lambda$ = 320 nm compared to reference devices with conventional aluminum doped zinc oxide (ZnO:Al) electrodes. The maximum responsivity of the multispectral photodetectors can be tuned in their wavelength from 320 nm to 510 nm by an external bias voltage, allowing single pixel detection of UV to visible light. Graphene electrodes further enable fully flexible diodes on polyimide substrates. Here, an upgrade from single to bilayer graphene boosts the maximum photoresponsivity from 134 mAW$^{-1}$ to 239 mAW$^{-1}$. Interference patterns that are present in conventional TCE devices are suppressed as a result of the atomically thin graphene electrodes. The proposed detectors may be of interest in fields of UV/VIS spectroscopy or for biomedical and life science applications, where the extension to the UV range can be essential.

Multispectral photodetectors (MS-PDs) provide an alternative solution to standard CCD/ CMOS sensors with color filters (Bayer pattern) or transverse field detectors [1] for the detection of specific spectral bands [2–4]. Unipolar [5,6], bipolar [3,7–9] and multi-terminal [10–12] hydrogenated amorphous silicon (a-Si:H) MS-PDs allow reconstructing color information in images without integrating color filters by utilizing a bias-tunable maximum of the spectral response (SR) [13]. The spectral response of state-of-the-art c-Si p-i-n- and p-n-junction PDs, in contrast, is limited to one single maximum [14], typically selected through color filters on individual pixels [15]. Alternatively, c-Si sensors with different spectral maxima can be stacked on top of each other [4,16], but this approach is limited to few different colors by the doping profiles [4, 17]. Nevertheless, such vertical sensor structures provide the opportunity to reduce the active sensor area and increase the spatial resolution of detector arrays significantly [18]. Filterless a-Si:H photodetectors can be grown vertically with graded doping profiles, such that high-energy photons are absorbed close to the sensor surface while low-energy photons will penetrate deeper into the sensor structure. Furthermore, the absorption coefficient of a-Si:H exceeds that of crystalline silicon (c-Si) in the visible spectrum between 400 nm up to 750 nm [19]. Charge carriers generated at different device depths can then be extracted by applying a negative bias voltage between the top and bottom electrode of the p-i-n structure, driving the device in the photo diode mode. All vertical multispectral detector concepts share the need for a transparent conductive electrode (TCE). Conventional transparent conductive oxides, e.g. aluminum doped zinc oxide (ZnO:Al), have been used in the past in a-Si:H MS-PDs and resulted in a tunable maximum spectral response from 420 nm to 630 nm [13]. However, the transmittance (T) of ZnO:Al electrodes decreases from 85% at visible wavelengths to approximately 20% at 350 nm, before it vanishes, limiting their applicability in the ultraviolet (UV) range [20]. Graphene, in contrast, shows high transmittance of approximately 97.7% over a large electromagnetic spectrum [21,22], which is considerably higher than that of other TCEs such as carbon nanotubes [23,24] and thin metallic films [25] or ITO [26]. In fact, Bae et al. demonstrated doped graphene monolayer transparent conductive electrodes with a sheet resistance down to ~125 Ω / □ with higher transmittance compared to ultrathin metals with a similar low sheet resistance[27]. Even though the transmittance of graphene is reduced for energies above 2.5 eV due to a van Hove singularity in its density of states [28], it remains above 90% for wavelengths > 200 nm (i.e. relevant to this work) and considerably higher than that of ZnO:Al. Graphene and other 2D materials lend themselves to vertical device concepts [29–32]. In addition, graphene is highly bendable / flexible, as opposed to brittle conductive oxides.

Here, we report on a-Si:H multispectral photodetectors with transparent conductive electrodes made of large area, chemical vapor deposited graphene. We demonstrate the superior efficiency of graphene-enabled MS diodes in the ultraviolet range. The entire fabrication process, including a-Si:H deposition and graphene transfer, is carried out at temperatures below 200°C. This ensures compatibility to CMOS technology, but also allows a variety of novel technologies such as Thin-Film-on-ASIC [33] or the deposition on organic [34] and other flexible substrates. In this work, we implement flexible a-Si:H detectors on Kapton polyimide substrates.

A schematic cross-section of an a-Si:H MS-PD as investigated in this work is shown in Figure 1a. The detector has a wide bandgap material (amorphous silicon carbide, a-SiC:H) close to its surface and a narrow bandgap material (amorphous silicon germanium, a-SiGe:H) near its backside. The schematic band diagram of such p-i-n diodes under reverse bias condition is shown in Fig. 1b. In these vertical devices, the charge carrier drift length $l_{drift}$ can be tuned with a bias voltage (i.e. electrical field E) to transport photon-generated charge carriers towards the contacts following equ. 1

$$l_{drift} = \mu\tau \cdot E \qquad (1)$$

where µ is the charge carrier mobility and τ is the charge carrier lifetime. The predominant location of charge generation in the device depends on the energy of photons: high energetic photons will generate electron-hole pairs near the top contact, while low energetic photons will penetrate deep into the device to generate charge carriers close to the bottom contact. If the carrier drift length exceeds the distance between the location of carrier generation and the electrical contacts, charge carriers are able to drift to the contact and contribute to a photocurrent. Tuning the carrier drift length with the bias voltage then allows collecting charge carriers from different depths in the device. Hence, one observes a shift of the maximum spectral response in the photocurrent as a function of bias voltage. The fabrication process for the devices is described in detail in the methods section.

A hybrid single layer graphene (SLG) / a-Si:H photodetector (Figure 2a) measured in the dark and under illumination with a tungsten halogen 3300 K white light source (8400 lux) exhibits a maximum dynamic range of 71.35 dB at a reverse bias voltage of $V_{bias}$ = -2.3 V (Figure 2c), defined as

$$DR = 20 \cdot log \frac{J_{8400\,lx}}{J_{dark}} \qquad (2)$$

The corresponding measurement of a PD with a ZnO:Al electrode (Figure 2b) reaches a maximum DR of 99.58 dB at $V_{bias}$ = -0.9 V (Figure 2d). This higher value can be attributed to a higher series resistance in the graphene/a-Si:H diodes: the electrode sheet resistances were measured independently on glass substrates to be 1.3 kΩ/□ for single layer graphene and 20 Ω/□ for the much thicker ZnO:Al. We note that the graphene electrode is smaller than the ZnO:Al electrode (0.81 mm² vs. 1 mm²) due to technological reasons: graphene had to be removed at the edges of the device to avoid short-circuiting. In addition, the J-/V characteristic of the reference device with ZnO:Al contact displays a built-in voltage of 0.4 V, while no considerable built-in voltage can be extracted from the J-/V curve of the graphene-contacted detector. We suppose that the low charge carrier density in graphene [41] is not sufficient to create a space-charge region under illumination.

Spectrally resolved photoresponse measurements were performed by selecting narrow wavelength bands (Δλ < 10 nm) from the light source with a monochromator in the spectral range from 300 nm to 800 nm. The setup was calibrated with a reference c-Si photodetector to achieve absolute spectral response (SR) values and the signals were acquired via lock-in technique. The absolute SR of both PD assemblies are shown for bias voltages from $V_{bias}$ = -1 V to $V_{bias}$ = -11 V in Figure and Figure 2f. These measurements can be used as indicators for the photon-/charge conversion efficiency [35]. The voltage-dependent SR of the SLG-PD reaches a maximum of 205 mAW$^{-1}$ at -11 V. The maximum external quantum efficiency (QE) of this device is 48.88%. QE is defined as

$$QE = \frac{hc}{q\lambda} \cdot SR \qquad (3)$$

where $h$ is the planck constant, $c$ is the speed of light, $q$ is the elemental charge of an electron and $\lambda$ is the wavelength of the light. The maximum SR of the sensor with the ZnO:Al electrode is 339 mAW$^{-1}$ (QE = 76.43%) at a bias of -11 V. Even though the performance of the conventional device appears to be superior due to lower series resistance, the advantage of graphene PD becomes apparent in the UV range (320 nm). This UV part of the spectrum is rather inaccessible with conventional TCEs. Here, the SLG device shows a UV response enhancement by a factor of 4.39 (SR = 118.16 mAW$^{-1}$, see Table 1) compared to the ZnO:Al reference device (SR = 26.94 mAW$^{-1}$).

Multispectral sensing occurs by reading out the wavelength of maximum response for a given bias voltage. The maximum SR for the SLG diode can be shifted from 380 nm at -1 V to 550 nm at -11 V, compared to a shift from 520 nm to 560 nm for the ZnO:Al electrode. The upper detection limit is almost identical in both cases (SLG: 550 nm, ZnO:Al: 560 nm), indicating that the light penetration and absorption in the deeper regions of the diode is similar for both electrodes. Since both detector types have been fabricated on the same chip at the same time, the difference in the low bias region, i.e. where absorption in the upper part of the diode dominates the photocurrent, must be due to the front contact. The precise origin of this difference is not entirely clear at the moment, but is not the main focus of this work. Nevertheless, a sharp drop of photocurrent below approximately 400 nm can be clearly observed in diodes with ZnO:Al electrodes, confirming previous results [36]. This drop is associated with the transmission properties of the ZnO:Al, which blocks UV light (Figure 3b).

An additional advantage of ultra-thin graphene electrodes are suppressed local SR maxima that are clearly visible in the ZnO:Al PD (here: at 380 nm, Figure 2f). These local maxima are due to interference patterns due to the refractive index mismatch of the ZnO:Al-/a-Si:H interface, as confirmed by simulations of the material stacks (Figure 2g, h). The optical simulations were performed with OPTICS which was developed for modeling the total transmittance, the total reflection and the absorption of each layer in a thin-film multilayer system for illumination at normal incidence. Iterative calculations are based on fundamental equations [37] describing the complex transmittance/reflections of electromagnetic waves at two thin film interfaces, including wavelength dependent phase shift and damping. The normalized absorption of the simulated pin-devices with graphene and ZnO:Al front electrodes take into account the wavelength dependent refractive indices and extinction coefficients. The material data for amorphous silicon [38], ZnO:Al [39] and graphene [40] was adapted from literature as well as partly taken from own measurements.

The absence of these local minima leads to a broader usable spectrum, because it enhances the signal linearity of graphene-based multispectral photodetectors, which is essential for further color corrections [18].

We further explore the opportunity to utilize graphene electrodes for flexible electronics on Kapton polyimide substrates. Gold bottom electrodes were deposited first, on which graded a-Si:H films were grown (see methods section for details). Single- and bilayer graphene (BLG) electrodes were then transferred as transparent electrodes, the latter in order to investigate the potential for optimizing diodes by reducing electrode sheet resistance. A photograph of a flexible MS a-Si:H PD with graphene electrodes on polyimide is shown in Figure 3a. The bilayer graphene exhibits a broadband transmittance of 96.5% (Figure 3b) and a sheet resistance of 600 Ω/□, measured on glass. ZnO:Al or other transparent conductive oxides are not suitable for this type of application because their brittleness limits mechanical bending of the devices [41].

A flexible multispectral diode with a single layer graphene electrode exhibits a maximum dynamic range of 44.89 dB at -3.3 V (Figure 3c). The device with a bilayer graphene electrode shows an almost six times higher current density in the J-/V curves (Figure 3d) under illumination than the SLG device, comparable to the performance of the rigid device. The maximum DR of the flexible bilayer PD is 59.79 dB at -3.6 V. Interestingly, we did not observe a shift of the threshold voltage (zero current) under illumination, a common feature in semiconductor PDs that depends on the bandgap of the semiconductor material and the fermi levels in the doped regions.

The absolute spectral response of the flexible PD with SLG electrode is shown in Figure 3e and exhibits bias-tunable maxima between 133.69 mAW$^{-1}$ at 510 nm and 67 mAW$^{-1}$ at 320 nm. In comparison, the flexible device with BLG electrode exhibits enhanced performance with maximum SRs of 238.57 mAW$^{-1}$ at 530 nm and 108.07 mAW$^{-1}$ at 320 nm, respectively (Figure 3f). The QE of the bilayer graphene device is 41.88% at 320 nm which is approximately four times higher than the rigid MS sensor with a ZnO:Al top contact (QE = 10.44%).

The main performance parameters of the detectors investigated in this work are summarized in Table 1.

The flexible devices were tested further on bending test fixtures (Figure 4a). J-/V curves were measured under strained conditions (Figure 4b, 4c). The radii of curvature were 25 mm and 6.35 mm, resulting in tensile strain of approximately 0.25% and 1%, respectively. Even after bending the sample 100 times with 0.25% strain and 10 times with 1% strain, the optoelectronic performance remained nearly constant. A similar detector with ZnO:Al, in contrast, changed its J-/V characteristics strongly when strained by 0.25% and failed completely at a tensile strain of ~1% (Figure 4c).

In conclusion, we presented the implementation of large-area graphene transparent conductive electrodes on rigid and flexible a-Si:H multispectral photodetectors. Graphene electrodes enable efficient broadband detection encompassing the UV range: the photoresponse response of 118.16 mAW$^{-1}$ (at λ = 320 nm) is four times higher for devices with graphene electrodes. The maximum responsivity of the detectors can be tuned from UV (320 nm) to VIS (510 nm) wavelengths. Flexible multispectral photodiodes on polyimide exhibit a maximum spectral response of 133.69 mAW$^{-1}$ with single layer graphene and 238.57 mAW$^{-1}$ with bilayer graphene electrodes. This demonstrates routes for optimizing the sheet resistance versus transparency of graphene TCEs. Graphene electrodes enable new types of flexible filterless multispectral photodetectors, which may ultimately be integrated in wearable systems. The broadband tunable spectral response offers prospects to apply the detectors in fields of fluorescence and spectrophotometric measurements, in chemical analysis, medical diagnostics or in colorimetric and multispectral imaging.

# Methods

Rigid MS devices were fabricated on 1 mm thick AF32 eco glass substrates and flexible graphene-/a-Si:H photodetectors were grown on 5 cm x 5 cm large and 125 µm thick Kapton polyimide films. A 220 nm ZnO:Al film was sputtered as a bottom electrode for the rigid devices. For the flexible diodes, a 180 nm gold bottom electrode was sputtered on polyimide. Then, a-Si:H thin film alloys were deposited on top of the bottom electrodes in a PE-CVD (Plasma Enhanced Chemical Vapor Deposition) high-vacuum multi-chamber system. For the deposition of the 4 nm n- and p-doped layer, a forward RF power of 8 W (13.56 MHz) and a low temperature of 200°C were chosen to minimize the thermal stress on the film. The films were p-doped with the addition of diborane ($B_2H_6$) and n-doped by phosphine ($PH_3$) in the regions sandwiching the ~500 nm intrinsic a-Si:H region (Figure 1a). To tune the bandgap from approximately 1.24 eV to 1.95 eV within the photo-active absorption layer, germane ($GeH_4$) and methane ($CH_4$) were added to the gas mixture of silane ($SiH_4$) and hydrogen ($H_2$) during the deposition [13]. CVD-grown SLG on copper [42] was transferred onto the a-Si:H stack via a PMMA assisted wet transfer [43]. This transfer process was repeated for the BLG devices. For the reference detector, a 220 nm thick ZnO:Al top electrode was sputtered (RF power: 150 W; chamber pressure: 8 mTorr) at 100°C onto the p-doped layer. The devices were structured by standard contact lithography. ZnO:Al was etched in 0.2% hydrochlorid acid and graphene was reactive ion etched (RIE) in $O_2$ plasma (80 sccm) at 50 W and a chamber pressure of 57 mTorr. The a-Si:H layers were structured by $SF_6$/Ar (100/50 sccm) RIE at 100 W and a chamber pressure of 100 mTorr. Electrical measurements were performed using a Keithley 4200-SCS parameter analyzer or lock-in amplifiers.

## Acknowledgements

We thank Deji Akinwande, Maruthi N. Yogeesh and Saungeun Park (University of Texas at Austin) as well as Olof Engström and Satender Katria (University of Siegen) for fruitful discussions. We thank Matthias Adlung (University of Siegen) for absorption spectra measurements. Funding by the German Research Foundation (DFG LE 2440/1-2, GRK 1564), the European Research Council through an ERC Starting Grant (InteGraDe, 307311) and the European Regional Funds (HEA2D) is gratefully acknowledged.

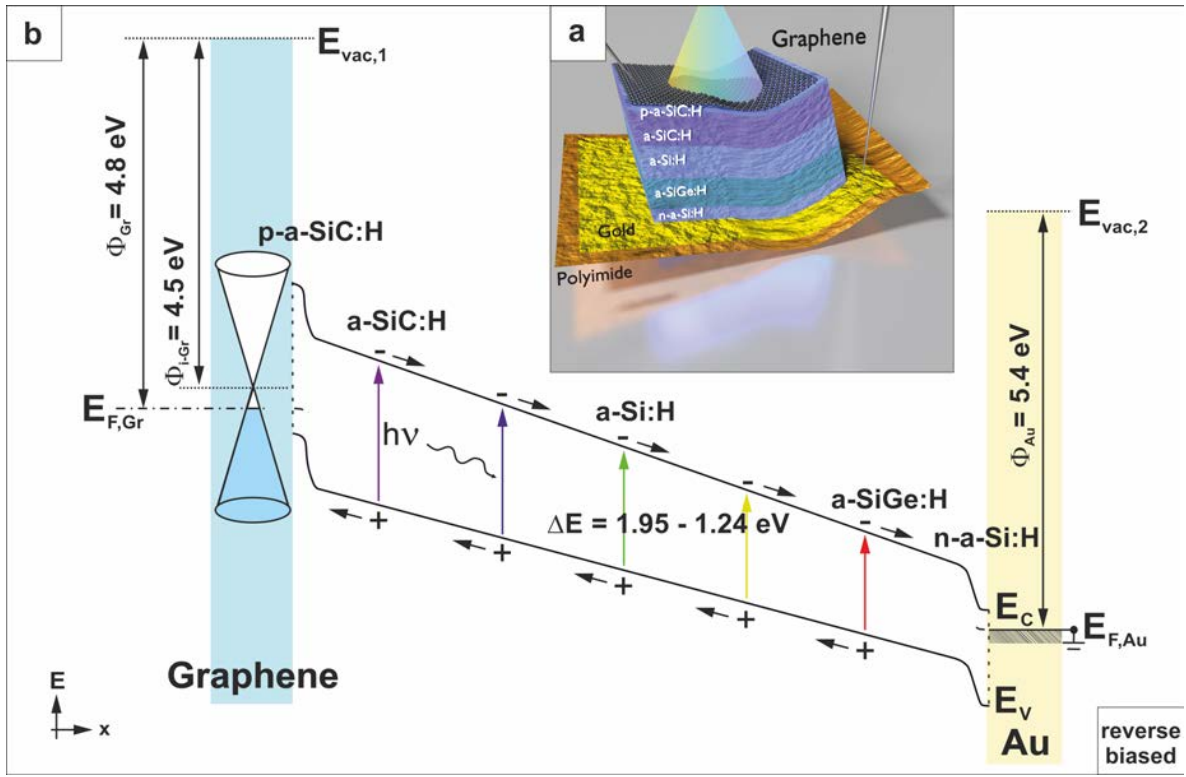

Figure 1: (a) Schematic cross section of a multispectral photodiode on polyimide substrate. (b) Energy-band diagram of a gradually doped a-Si:H / graphene multispectral photodetector in reverse bias conditions.

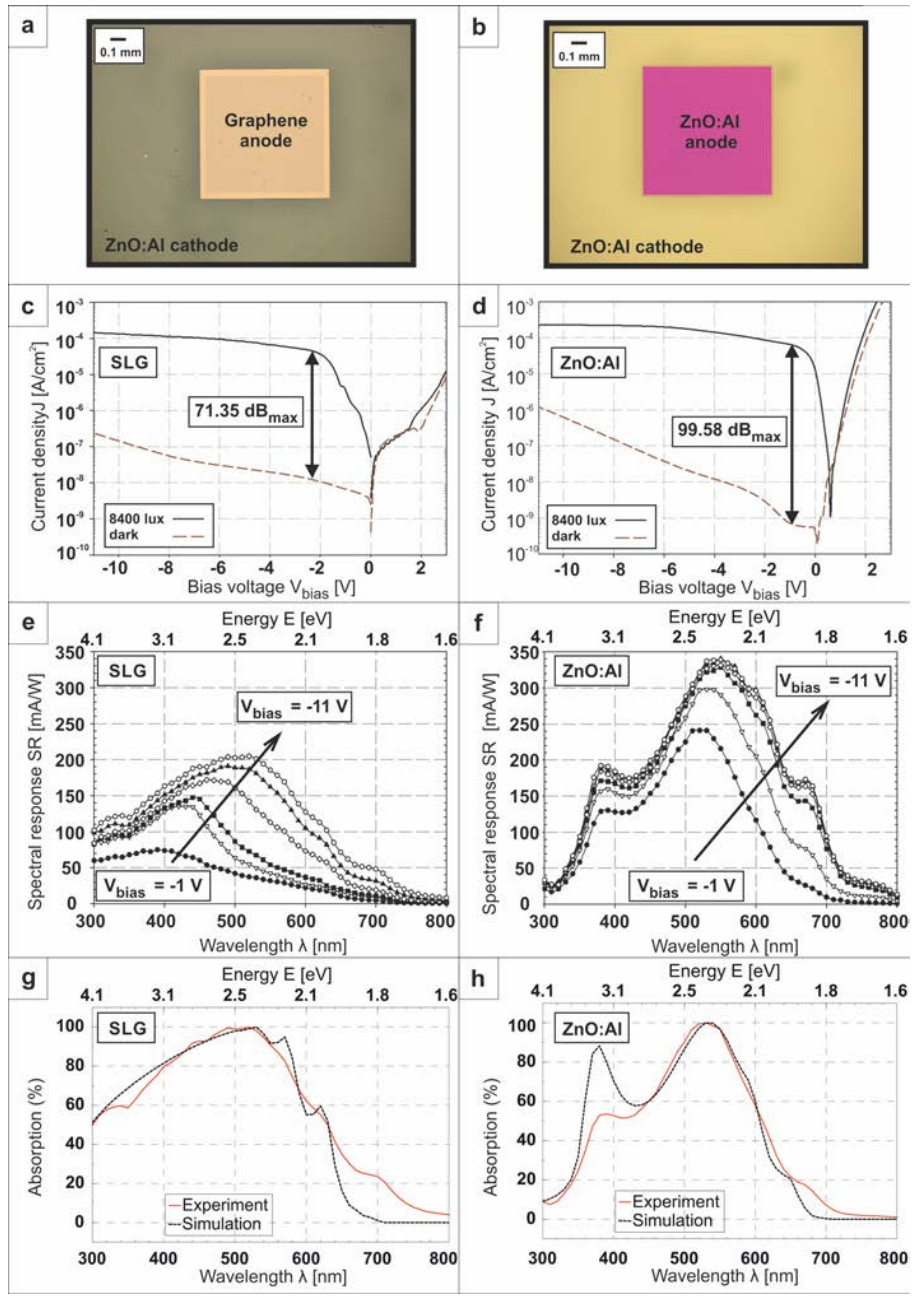

Figure 2: (a) and (b) top views of the SLG/a-Si:H and ZnO:Al/a-Si:H photodiode assemblies, (c) and (d) electrical characteristics under dark conditions and under white-light illumination, (e) and (f) absolute SR for different bias voltages. (g) and (h) optical simulations of the normalized absorption of amorphous silicon PDs with SLG and 220 nm ZnO:Al anodes.

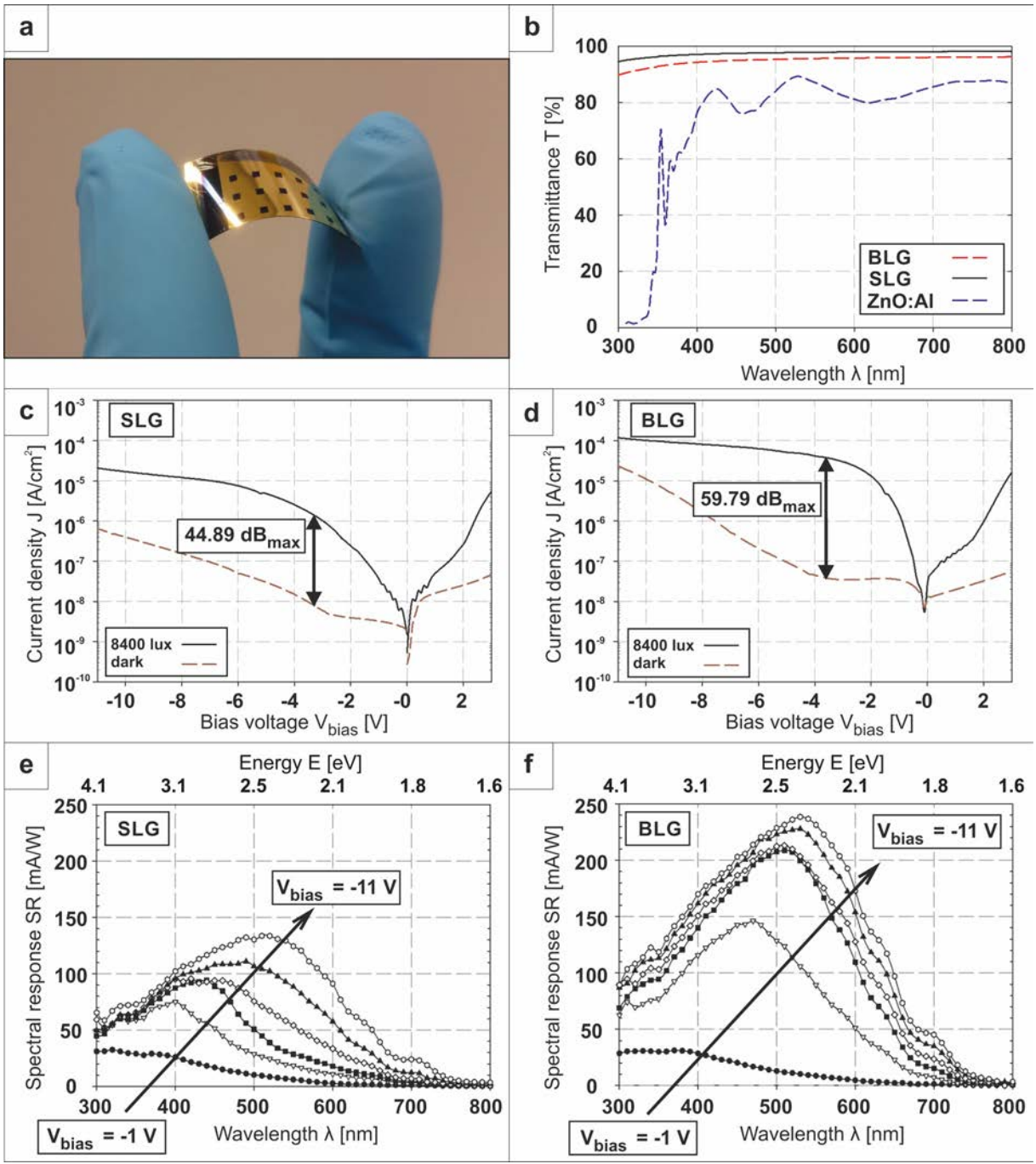

Figure 3: (a) Photograph of the SLG/a-Si:H MS-PDs on a flexible PI substrate, (b) measured transmittance of SLG, BLG and ZnO:Al. Electrical characteristics of (c) SLG and (d) BLG PDs under dark conditions and under white-light illumination. Absolute spectral response of (e) SLG and (f) BLG MS-PDs.

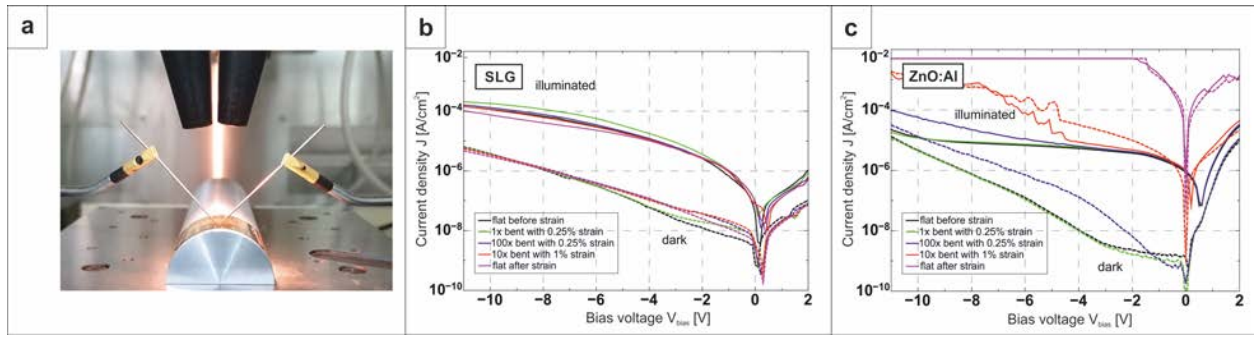

Figure 4: (a) Photograph of a flexible SLG/a-Si:H MS-PD on a 25 mm radius fixture aperture for electro-optical characterizations under bent conditions. (b) Comparison of photocurrent measurements in initial condition, during and after bending of a PD with SLG and (c) ZnO:Al anode.

Table 1: Electrical and optical characteristics of the a-Si:H multispectral photodetectors investigated in this work.

| Multispectral Photodetector | Max. DR [dB] | Max. SR at -11 V | UV response enhancement ($\lambda$ = 320 nm) | Top contact | | |
|---|---|---|---|---|---|---|
| | | | | $R_s$ [$\Omega/\square$] | T [%] | area |
| **ZnO:Al/a-Si:H** | 99.58 (at -0.9 V) | 339 mAW$^{-1}$ ($\lambda$ = 550 nm) | 1* (SR = 26.94 mAW$^{-1}$) | 20 | ≈85 | 1 mm² |
| **SLG/a-Si:H** | 71.35 (at -2.3 V) | 204.83 mAW$^{-1}$ ($\lambda$ = 520 nm) | 4.39 (SR = 118.16 mAW$^{-1}$) | 1.3 k | 98.4 | 0.81 mm² |
| **Flexible SLG/a-Si:H** | 44.89 (at -3.3 V) | 133.69 mAW$^{-1}$ ($\lambda$ = 510 nm) | 2.49 (SR = 67 mAW$^{-1}$) | 1.3 k | 98.4 | 0.81 mm² |
| **Flexible BLG/a-Si:H** | 59.80 (at -3.6 V) | 238.57 mAW$^{-1}$ ($\lambda$ = 530 nm) | 4.01 (SR = 108.07 mAW$^{-1}$) | 600 | 96.5 | 0.81 mm² |

* reference detector = ZnO:Al front electrode